\documentclass[epj]{svjour}

\RequirePackage{graphicx}

\usepackage{cite} 
\usepackage{amsmath}
\usepackage{hyperref}
\usepackage{amsmath,amssymb}
\usepackage{xcolor}
\usepackage{graphics}

\newcommand{\be}{\begin{equation}}
\newcommand{\en}{\end{equation}}
\newcommand{\bea}{\begin{eqnarray}}
\newcommand{\ena}{\end{eqnarray}}

\begin{document}

\title{Quasinormal modes of five-dimensional black holes in non-commutative geometry}

\author{
Grigoris Panotopoulos \inst{1} 
\thanks{E-mail: \href{mailto:grigorios.panotopoulos@tecnico.ulisboa.pt}{\nolinkurl{grigorios.panotopoulos@tecnico.ulisboa.pt}} }
\and
\'Angel Rinc\'on \inst{2}
\thanks{E-mail: \href{mailto:angel.rincon@pucv.cl}{\nolinkurl{angel.rincon@pucv.cl}} }
}                     
%
%
\institute{ 
Centro de Astrof{\'i}sica e Gravita{\c c}{\~a}o, Departamento de F{\'i}sica, Instituto Superior T\'ecnico-IST,
\\
Universidade de Lisboa-UL, Av. Rovisco Pais, 1049-001 Lisboa, Portugal
\and
Instituto de F{\'i}sica, Pontificia Universidad Cat\'olica de Valpara{\'i}so, Avenida Brasil 2950, Casilla 4059, Valpara{\'i}so, Chile
}

\date{Received: date / Revised version: date}
%
\abstract{
We compute the spectrum of quasinormal frequencies of five-dimensional black holes obtained in noncommutative geometry. In particular, we study scalar perturbations of a massive scalar field adopting the 6th order WKB approximation. We investigate in detail the impact of the mass of the scalar field, the angular degree and the overtone number on the spectrum. All modes are found to be stable.
}

\maketitle

\section{Introduction}

The singularity at the center of black holes (BHs) in Einstein's General Relativity (GR) \cite{GR} is hidden by an event horizon, and therefore it has no effect on the outside region, where Physics is well-behaved. The existence of singularities, however, indicate the breakdown of General Relativity, and so attempts are made to obtain regular BH solutions, such as the solution obtained for the first time by Bardeen \cite{Bardeen}, see also \cite{borde}. One way to achieve that is to assume appropriate non-linear electromagnetic sources, which in the weak field limit are reduced to the standard Maxwell's linear theory. This approach allows us to generate a new class of solutions to Einstein's field equations \cite{beato1,beato2,beato3,bronnikov,dymnikova,hayward,vagenas1,vagenas2}, which on the one hand have a horizon, and on the other hand their curvature invariants, such as the Ricci scalar $R$, are regular everywhere, as opposed to the standard Reissner-Nordstr{\"o}m solution \cite{RN}. Regular BHs may help us understand the final states of gravitational collapse \cite{ellis,extra}, which is not possible when singularities are present.

Another way to obtain regular black hole solutions is to assume a noncommutative (NC) spacetime \cite{Connes1,Connes2,Connes3}. Noncommutativity lies at the heart of quantum physics through the uncertainty principle, and it can be linked to Superstring Theory \cite{Douglas,Seiberg}, which is a consistent theory of gravity, and which is characterized by several remarkable properties. To mention a few, it is compatible with both relativity and quantum physics, it is finite, and it contains General Relativity together with the gauge interactions of the Standard Model of Particle Physics (for standard textbooks see \cite{ST1,ST2}). Superstring theory has put forward the idea that extra spacelike dimensions may exist, not only because the theory itself is formulated in 10 dimensions, but also because of the brane-world scenario as well as the AdS/CFT correspondence. In the brane-world scenario it is assumed that the Standard Model must be confined on a three-dimensional hypersurface (the brane), while at the same time there are additional dimensions transverse to the brane, and gravitons can freely propagate into the bulk \cite{ADD,RS1,RS2,DGP}. In the AdS/CFT correspondence \cite{maldacena}, and more generically in the gauge-gravity duality \cite{klebanov}, one can understand strongly coupled field theories in d dimensions by studying a weakly coupled gravitational theory in d+1 dimensions \cite{guide}.

When BHs are perturbed the geometry of spacetime undergoes dumped oscillations due to the emission of gravitational waves. The so called quasinormal modes (QNMs) are complex frequencies that encode the information on how black holes relax after the perturbation has ceased to act on them, and they enter into the ring down phase of a black hole merger after the formation of the single distorted object. The work of \cite{wheeler} marked the birth of BH perturbations, it was later extended by \cite{zerilli1,zerilli2,zerilli3,moncrief,teukolsky}, while a comprehensive overview of BH perturbations is summarized in Chandrasekhar's monograph \cite{monograph}. Although perturbations of black holes is an old subject, after the LIGO historical direct detection of gravitational waves \cite{ligo1,ligo2,ligo3,ligo4,ligo5}, which has provided us with the strongest evidence so far that black holes do exist in Nature, there is nowadays a renewed interest in studying the QNMs of black holes and their alternatives, such as exotic compact objects \cite{pani}. For a review on the subject see \cite{review1}, and for a more recent ones \cite{review2,review3}.

Over the years the computation of the QNMs of higher-dimensional BHs has attracted a lot of attention for several reasons, namely i) the study of features of higher-dimensional GR \cite{extra1,extra2}, ii) the analysis of the physical implications of the brane-world scenario \cite{kanti}, and iii) the understanding of thermodynamic properties of BHs in Loop Quantum Gravity \cite{extra3,extra4}. Given the interest in Gravitational Wave Astronomy and in QNMs of black holes, it would be interesting to see what kind of QN spectra are expected from regular BHs within the framework of noncommutative geometry in various space time dimensions (see e.g. \cite{Lemos_New,GPAR,churilova} for QNMs of regular charged black holes with non-linear electrodynamic sources). In particular, it was shown that the gravitational wave signal from the event GW150914, detected by the LIGO and Virgo collaborations \cite{ligo1}, could be used to obtain a bound on the scale of quantum fuzziness of noncommutative space-time \cite{Kobakhidze:2016cqh}. QNMs of the BTZ black hole as well as four-dimensional black holes in noncommutative geometry have been studied in \cite{work1,work2,work3,work4,work5,RefExtra1}, while QNMs and gravitational radiation of standard higher-dimensional BHs have been studied in \cite{5D1,5D2,molina,master,5D3,5D4,5D5,ortega1,ortega2,ortega3,ortega4,panotopoulos}.

It is the goal of the present article to compute the QNMs of scalar perturbations of five-dimensional noncommutative black holes. Our work is organized as follows: After this Introduction, we present the wave equation for scalar perturbations in Section \ref{WE_QNM}. In the third Section we compute the QNMs of the black holes in the WKB approximation and we discuss our results. Finally, we conclude our work in Section \ref{Conclusions}. We use natural units such that $c = G = 1$ and metric signature $(-, +, +, +, +)$.

\section{Scalar perturbations of NC black holes}

\subsection{Noncommutative black hole in five dimensions}

The key feature of noncommutative geometry is the discretization of spacetime. The realization of such an idea becomes into noncommuting operators on a D-brane \cite{Smailagic:2003yb,NCBH1}. To be more precise
noncommutativity of spacetime is encoded into the commutator
\be \label{comm}
[x^\mu, x^\nu] = i \Theta^{\mu \nu}
\en
where $\Theta^{\mu \nu}$ is an anti-symmetric matrix. Without loss of generality it may be taken to have the Jordan form
\be
\Theta^{\mu \nu} = \Theta \text{diag}(\epsilon_{ij},\epsilon_{ij},...)
\en
where $\Theta$ is the noncommutative parameter, and $\epsilon_{ij}$ is the 2-dimensional anti-symmetric matrix
\be
\epsilon_{ij} = ((0,1),(1,0)) 
\en
It has been shown that to obtain noncommutative black hole solutions, we can still use the usual Einstein's equations
\be
G_{\mu \nu} = 8 \pi T_{\mu \nu}
\en
using an appropriate stress-energy tensor for matter \cite{NCBH1,NCBH2}. In particular, noncommutativity eliminates point-like structures in favour of smeared objects \cite{NCBH1,Ivan}. To obtain spherically symmetric black hole solutions, we employ the coordinate system $(t,r,\theta,\phi,\psi)$
and we make the following ansatz for the metric tensor
\be
\mathrm{d}s^2 = - f(r) \mathrm{d}t^2 + f(r)^{-1} \mathrm{d}r^2 + r^2 \mathrm{d} \Omega_3^2
\en
where $\mathrm{d} \Omega_3^2$ is the line element of the unit three-dimensional sphere given by \cite{EGB1,EGB2}
\be
\mathrm{d} \Omega_3^2 = \mathrm{d} \theta^2 + \sin^2 \theta \mathrm{d} \phi^2 + \sin^2 \theta \sin^2 \phi \mathrm{d}\psi^2
\en
one has to assume a stress-energy momentum for matter of the form \cite{NCBH1,NCBH2}
\be
\mathrm{d}s^2 = \text{diag}(-\rho, p_r, p_t, p_t, p_t)
\en
where the radial pressure $p_r$ and the tangential pressure $p_t$ are given in terms of the energy density $\rho$ as follows \cite{NCBH2}
\be
p_r = - \rho
\en
\be
p_t = -\rho - \frac{1}{3}r \rho' = \left(1-\frac{r^2}{6 \Theta}\right)  \rho
\en
while the energy density is given by
\be
\rho = \frac{M}{(4 \pi \Theta)^2} \exp\left(-{\frac{r^2}{4 \Theta}}\right)
\en
so that the mass of the black hole $M$ is the total mass of the matter distribution
\be
M = \int \mathrm{d}^4 \vec{r} \: \rho(r)
\en
where we have made use of the following formula \cite{5D1,5D2}
\begin{equation}
\Omega_{D-2} = \frac{2 \pi^{(D-1)/2}}{\Gamma\left( \frac{D-1}{2} \right)}
\end{equation}
for the surface of the unit $(D-1)$-dimensional sphere.
Using the tt Einstein's equation one can determine the unknown metric function $f(r)$, which is found to be \cite{NCBH2}
\be
f(r) = 1 - \frac{8 M}{3 \pi r^2} \: \gamma \left(2,\frac{r^2}{4 \Theta}\right)
\en
where $\gamma(a,z)$ is the lower incomplete Gamma function defined by
\be
\gamma(a,z) \equiv \int_0^z \mathrm{d}t e^{-t} t^{a-1}
\en
It can be easily verified that the rest of the equations as well as the trace equation $R=-(16 \pi T)/3$, with $R$ being the Ricci scalar and $T=T_\mu^\mu$ is the trace of the stress-energy tensor, are satisfied too. 
Clearly, when $\Theta \rightarrow 0$, we recover the standard five-dimensional Schwarzschild solution \cite{Tangherlini}. 
Notice that the Ricci scalar is found to be
\be
R = - \frac{M (r^2 - 10 \Theta)}{6 \pi \Theta^3} \exp \left(-\frac{r^2}{4 \Theta}\right)
\en
which is clearly non-singular. 

The lapse function $f(r)$ determines the horizon of the black hole requiring that $f(r_H)=0$. It is not possible to obtain an expression in a closed form, but we can still express the mass in terms of the event horizon as follows:
\begin{align}
M = \frac{3 \pi r_H^2}{8} \left[ \gamma \left(2,\frac{r_H^2}{4 \Theta} \right) \right] ^{-1}.
\end{align}
Thus, depending on the mass of the black hole $M$ for a given $\Theta$ there are 3 distinct cases: 
i) There is no horizon for small masses, $M < M_0$, 
ii) there is an inner horizon $r_-$ and an event horizon $r_H$ for large black hole masses, $M > M_0$, and
iii) there is single (event) horizon for the critical mass, $M=M_0$. This case corresponds to an extremal black hole, where $r_-=r_H$.

The metric function for all 3 cases is shown in the left panel of Fig.~\ref{fig:potentials} for $M=1$ 
and $\Theta=0.04,0.0634,0.08$.

\subsection{Perturbations for a test massive scalar field}\label{WE_QNM}

Before we consider the propagation of a scalar field in a curved spacetime, we need to briefly report on the effect of noncommutativity on a scalar field theory in flat spacetime. In this work we shall be interested in a free massive real scalar field described by the Lagrangian density
\be
\mathcal{L} = \frac{1}{2} (\partial \phi)^2 - \frac{1}{2}m^2 \phi^2
\en
without any interaction terms, with $m$ being the mass of the field.
In noncommutative field theories deformations are induced via the Moyal product, or "star product", which replaces the usual product, and which is defined by \cite{Moyal}
%
%
\be
(f \star g)(x) = f(x) \exp\left( \frac{1}{2} i \Theta^{\mu \nu} \overleftarrow{\partial}_\mu \overrightarrow{\partial}_\nu \right) g(x)
\en

for any two functions $f,g$ of the spacetime point $x$. However, quadratic terms in the action are the same both in the usual and in the case of commutativity \cite{lectures}, and therefore deformations are expected through interactions only. 

\begin{figure*}[ht]
\centering
\includegraphics[width=0.32\textwidth]{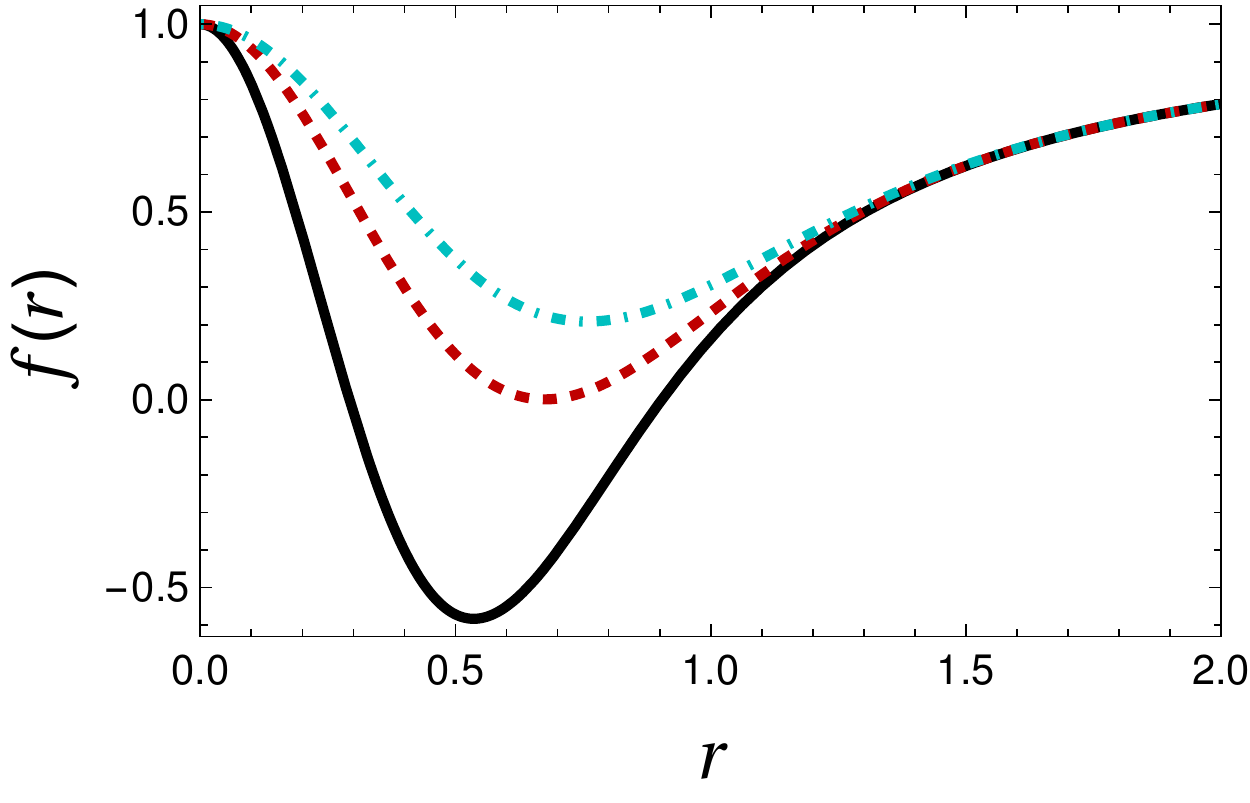}  \ 
\includegraphics[width=0.32\textwidth]{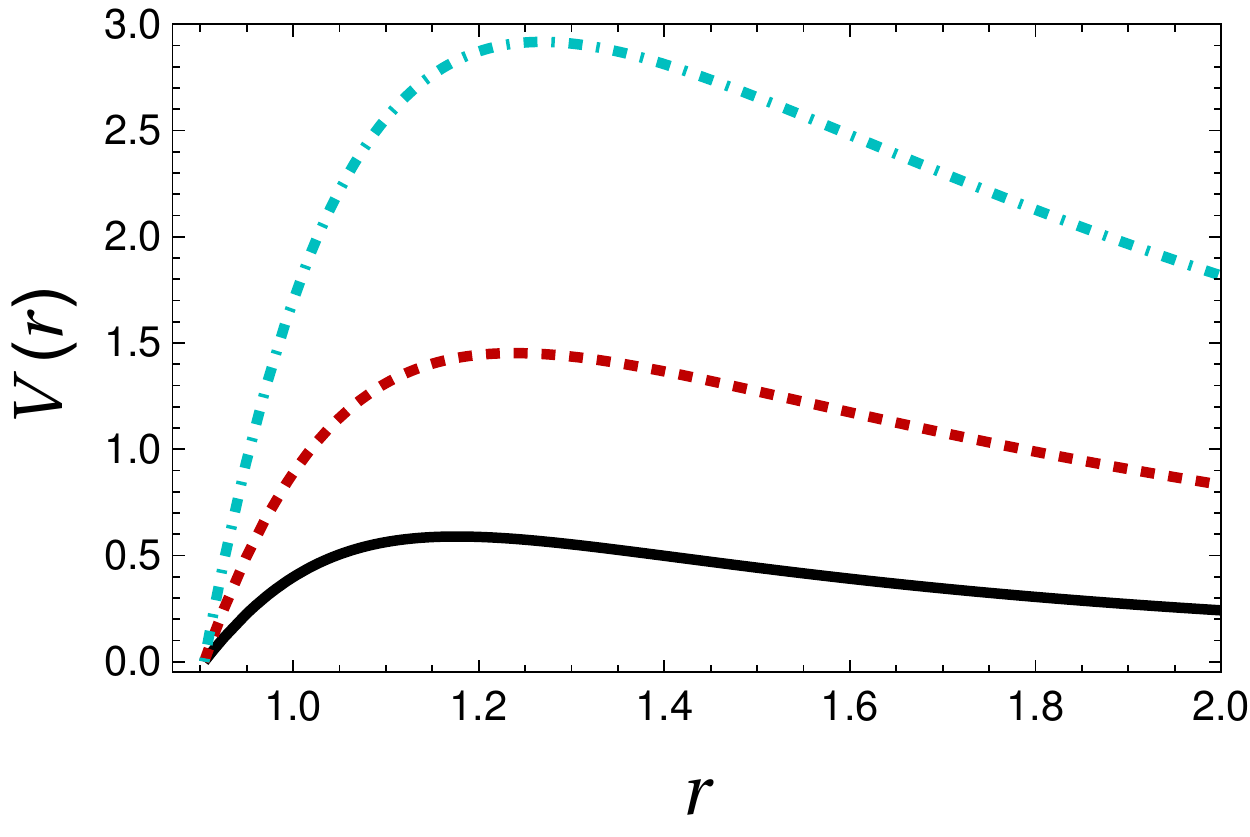}   	\
\includegraphics[width=0.32\textwidth]{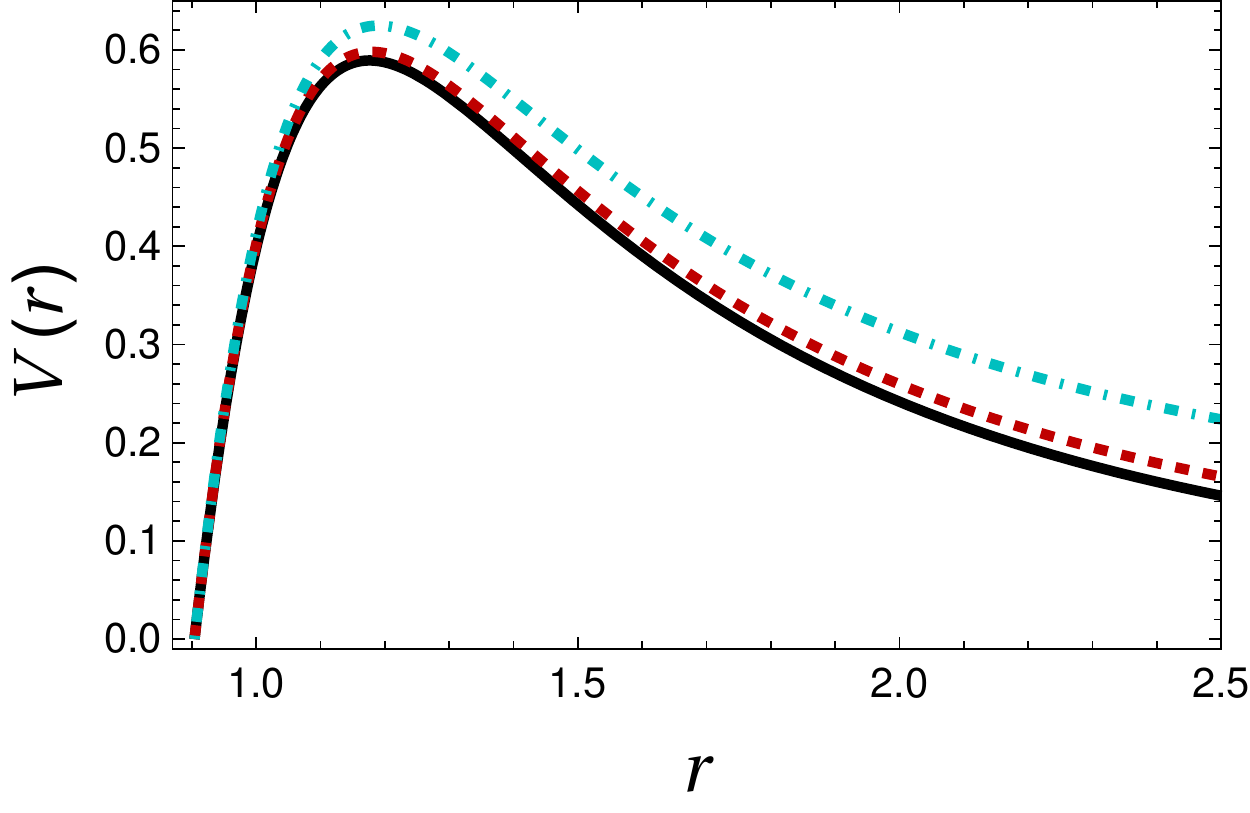}   	\
\caption{
{\bf{Left:}}
Lapse function vs radial coordinate for $M=1$ and $\theta=0.04$ (there is an event horizon as well as an inner horizon), $\theta=0.08$ (there are no horizons), and $\theta=0.0634$ (critical case).
{\bf{Middle:}}
Effective potential vs radial coordinate for $M=1, \theta=0.04$ for massless scalar field and $l=0,1,2$ from bottom to top.
{\bf{Right:}}
Effective potential vs radial coordinate for $M=1, \theta=0.04$ for $l=0$ and $\mu=0,0.15,0.3$ from bottom to top. 
}
\label{fig:potentials}
\end{figure*}

We perturb the black hole with a probe minimally coupled massive scalar field with equation of motion
\begin{equation}
\frac{1}{\sqrt{-g}} \partial_\mu (\sqrt{-g} g^{\mu \nu} \partial_\nu) \Phi = \mu^2 \Phi
\end{equation}
where $\mu$ is the mass of the test scalar field, and we consider the propagation of the test scalar field in the fixed gravitational background of the previous subsection. We separate variables making the standard ansatz
\begin{equation}\label{separable}
\Phi(t,r,\theta, \phi,\psi) = e^{-i \omega t} \frac{\Psi(r)}{r^{3/2}} \tilde{Y}_l (\Omega)
\end{equation}
with $\tilde{Y}_l (\Omega)$ being the higher-dimensional generalization of the usual spherical harmonics depending on the angular coordinates \cite{book}, and we obtain a Schr{\"o}dinger-like equation of the form
\begin{equation}
\frac{\mathrm{d}^2 \Psi}{\mathrm{d}x^2} + (\omega^2 - V(x)) \Psi = 0
\end{equation}
with $x$ being the so-called tortoise coordinate
\begin{equation}
x  =  \int \frac{\mathrm{d}r}{f(r)}
\end{equation}
while the effective potential is given by the expression \cite{5D1,5D2}
\begin{equation}
V(r) = f(r) \: \left( \mu^2 + \frac{l (l+2)}{r^2}+\frac{3 f'(r)}{2 r} +\frac{3 f(r)}{4 r^2} \right)
\end{equation}
where $l (l+2)$ is the eigenvalue of the Laplace operator on the $S^3$ hypersurface, and the prime denotes differentiation with respect to $r$.
The effective potential as a function of the radial coordinate can be seen in the middle (for $l=0,1,2$ from bottom to top) and right panels (for $\mu=0,0.15,0.3$ from bottom to top) of Fig. \eqref{fig:potentials}.

For asymptotically flat spacetimes the Schr{\"o}dinger-like equation is supplemented by the boundary conditions at the horizon and at infinity \cite{valeria}

\begin{equation}
\Psi(x) \rightarrow
\left\{
\begin{array}{lcl}
A e^{-i \omega x} & \mbox{ if } & x \rightarrow - \infty \\
&
&
\\
 C e^{i \omega x}  & \mbox{ if } & x \rightarrow + \infty
\end{array}
\right.
\end{equation}
where $A, C$ are arbitrary coefficients. The purely ingoing wave physically means that nothing can escape from the horizon, while the purely outgoing wave corresponds to the requirement that no radiation is incoming from infinity \cite{valeria}. The quasinormal condition allows us to obtain an infinite set of discrete complex numbers called the quasinormal frequencies of the black hole. Given the time dependence of the scalar field, $\sim e^{-i \omega t}$, the mode is unstable (exponential growth) when $\omega_I > 0$ and stable (exponential decay) when $\omega_I < 0$. In the latter case the real part determines the frequency of the oscillation, $\omega_R/(2 \pi)$, while the inverse of $|\omega_I|$ determines the dumping time, $t_D^{-1}=|\omega_I|$.


\begin{figure*}[ht]
\centering
\includegraphics[width=0.49\textwidth]{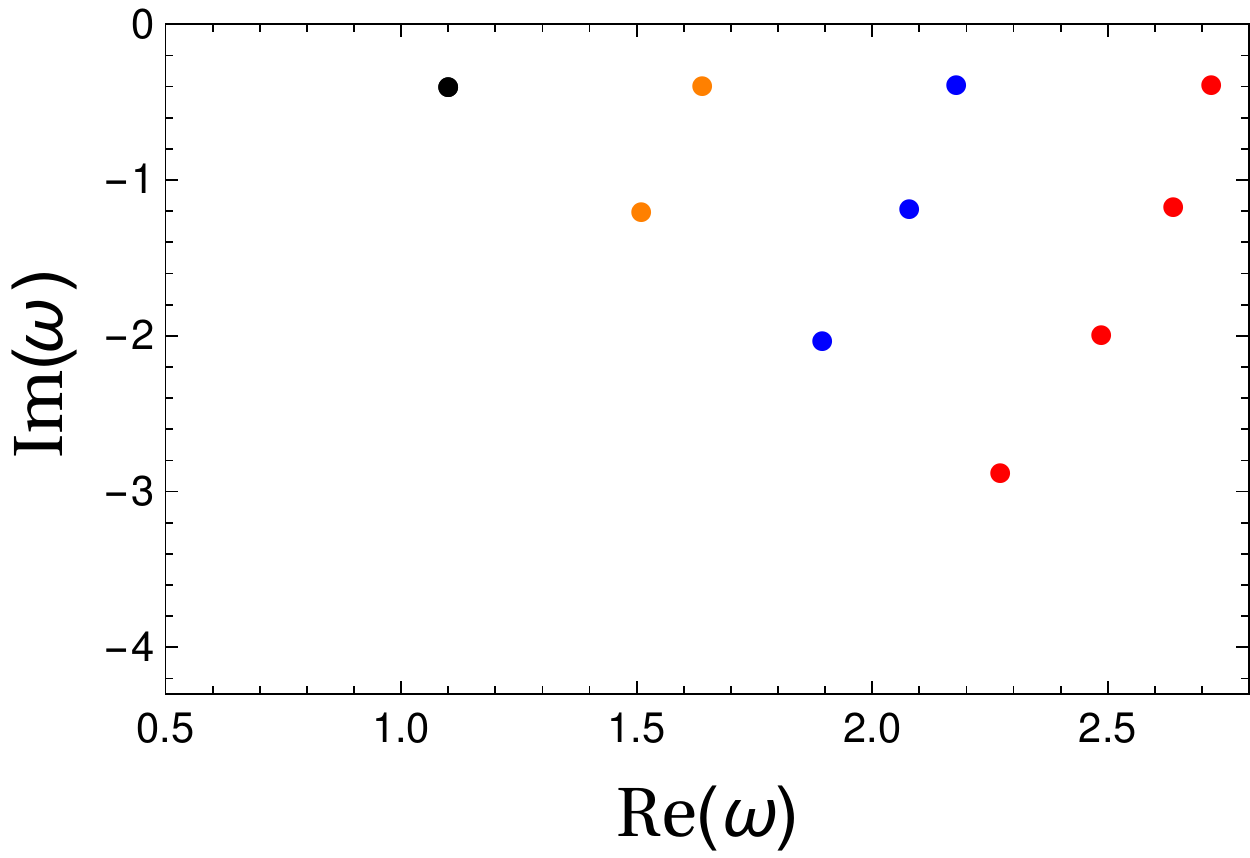}  \ 
\includegraphics[width=0.49\textwidth]{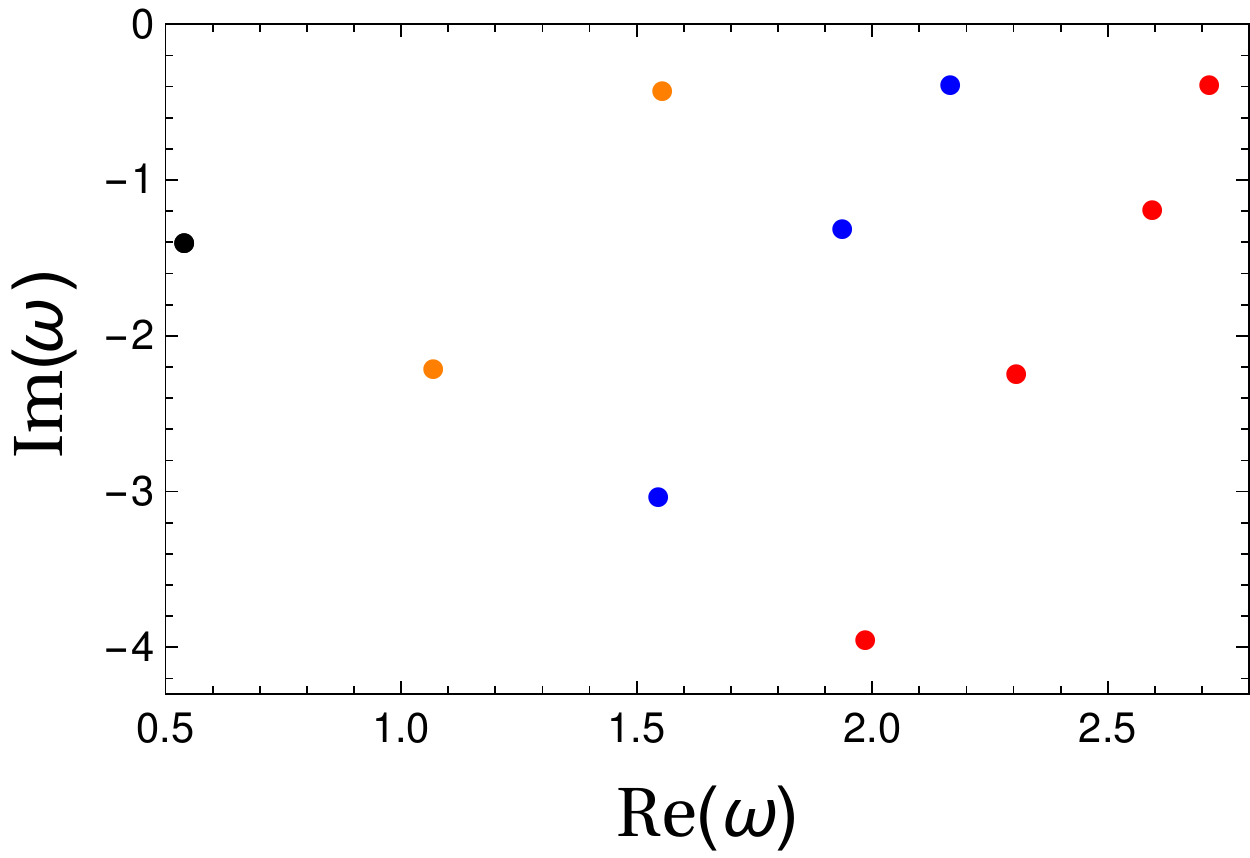}  \ 
\caption{
Imaginary part vs real part of the QN frequencies computed for a massless test scalar field, $M=1, \theta=0.04$ (right panel), and $l=1$ (black), $l=2$ (orange), $l=3$ (blue) and $l=4$ (red). For comparison reasons we show the modes of the standard 5D black hole (left panel).
}
\label{fig:spectrum1}
\end{figure*}

\begin{figure*}[ht]
\centering
\includegraphics[width=0.49\textwidth]{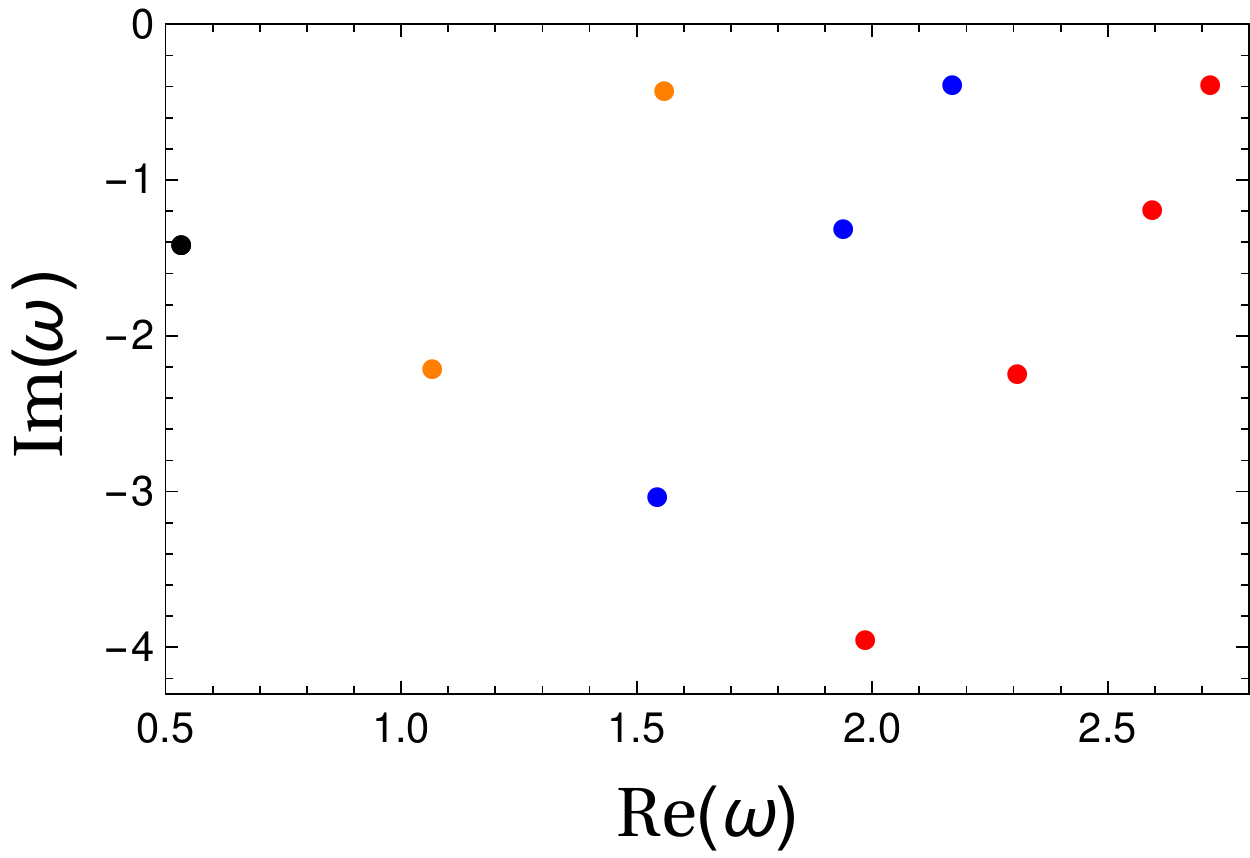}   \
\includegraphics[width=0.49\textwidth]{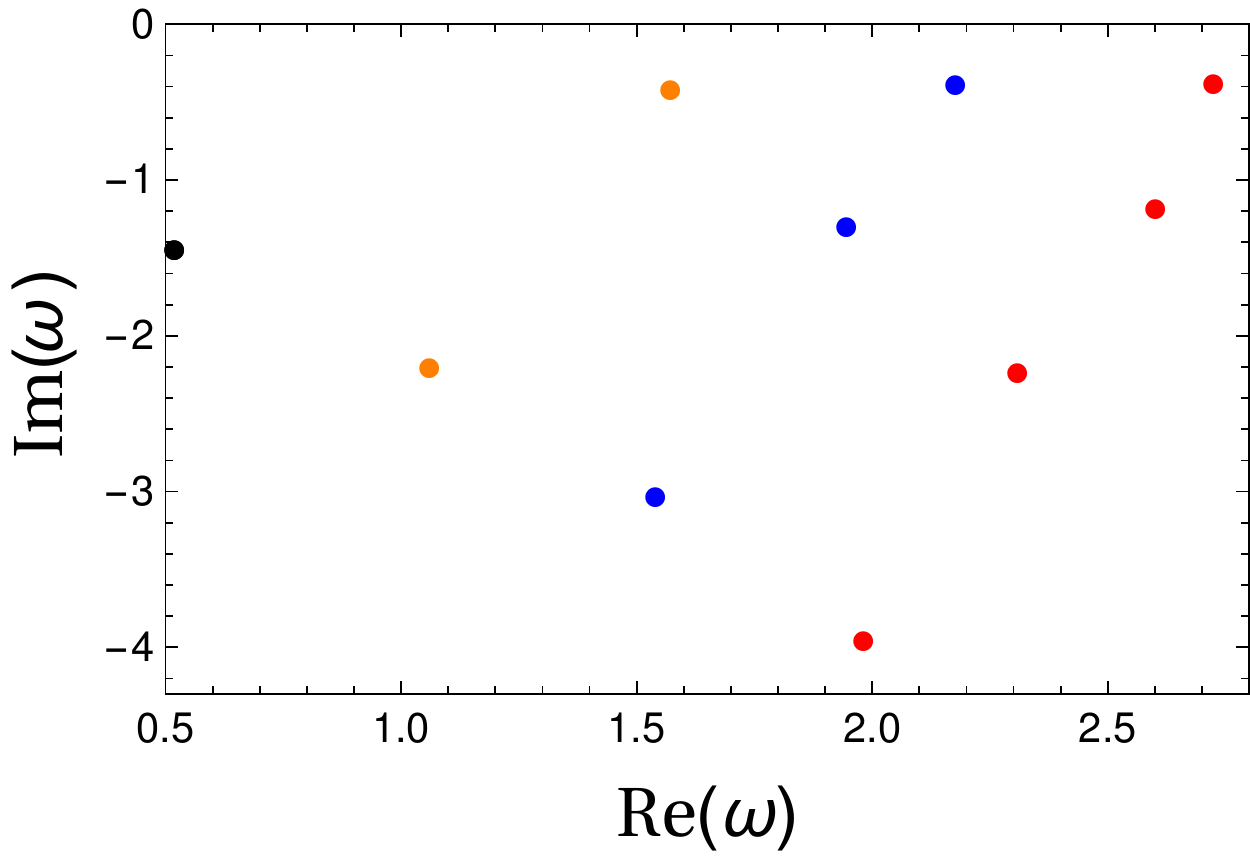}   \
\caption{
Imaginary part vs real part of the QN frequencies computed for a massive test scalar field for $M=1, \theta=0.04$, and $l=1$ (black), $l=2$ (orange), $l=3$ (blue) and $l=4$ (red). From left to right we have the cases $\mu = 0.15$ and $\mu=0.3$, respectively.
}
\label{fig:spectrum2}
\end{figure*}

\begin{figure*}[ht]
\centering
\includegraphics[width=0.49\textwidth]{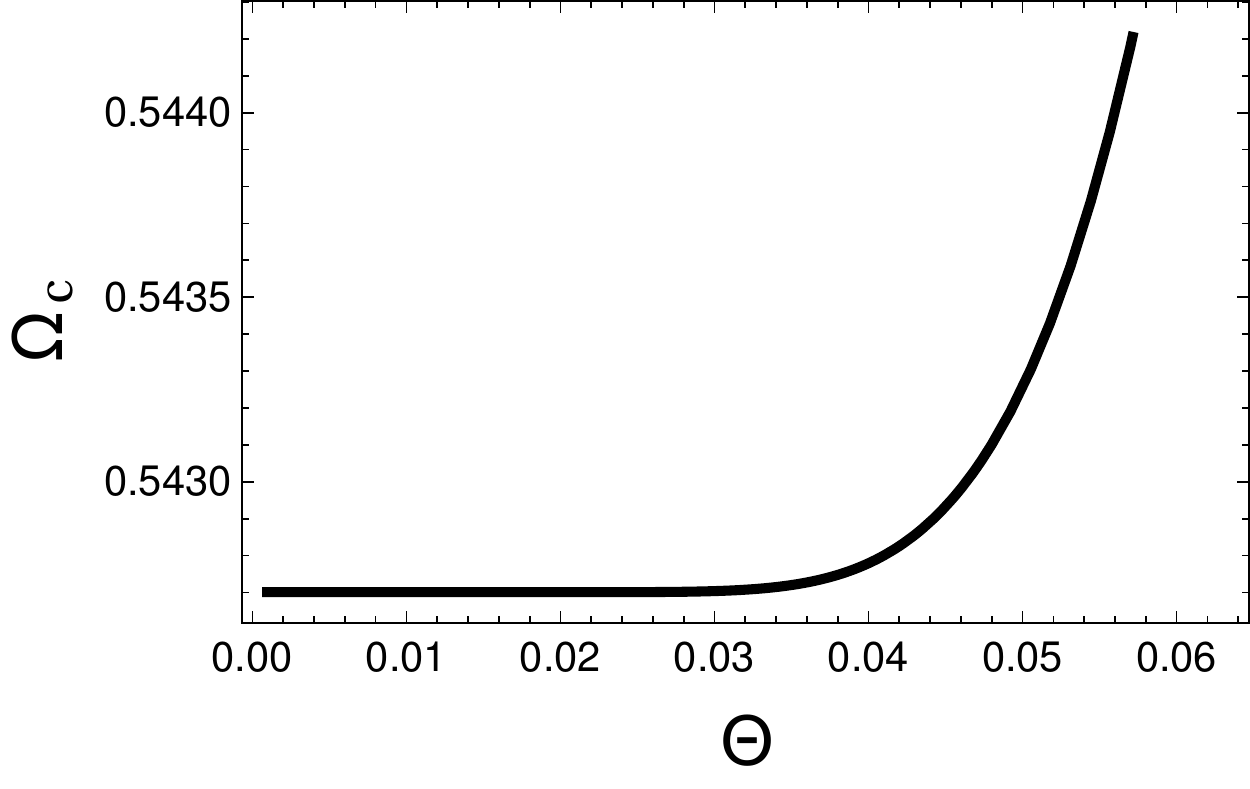}   \
\includegraphics[width=0.49\textwidth]{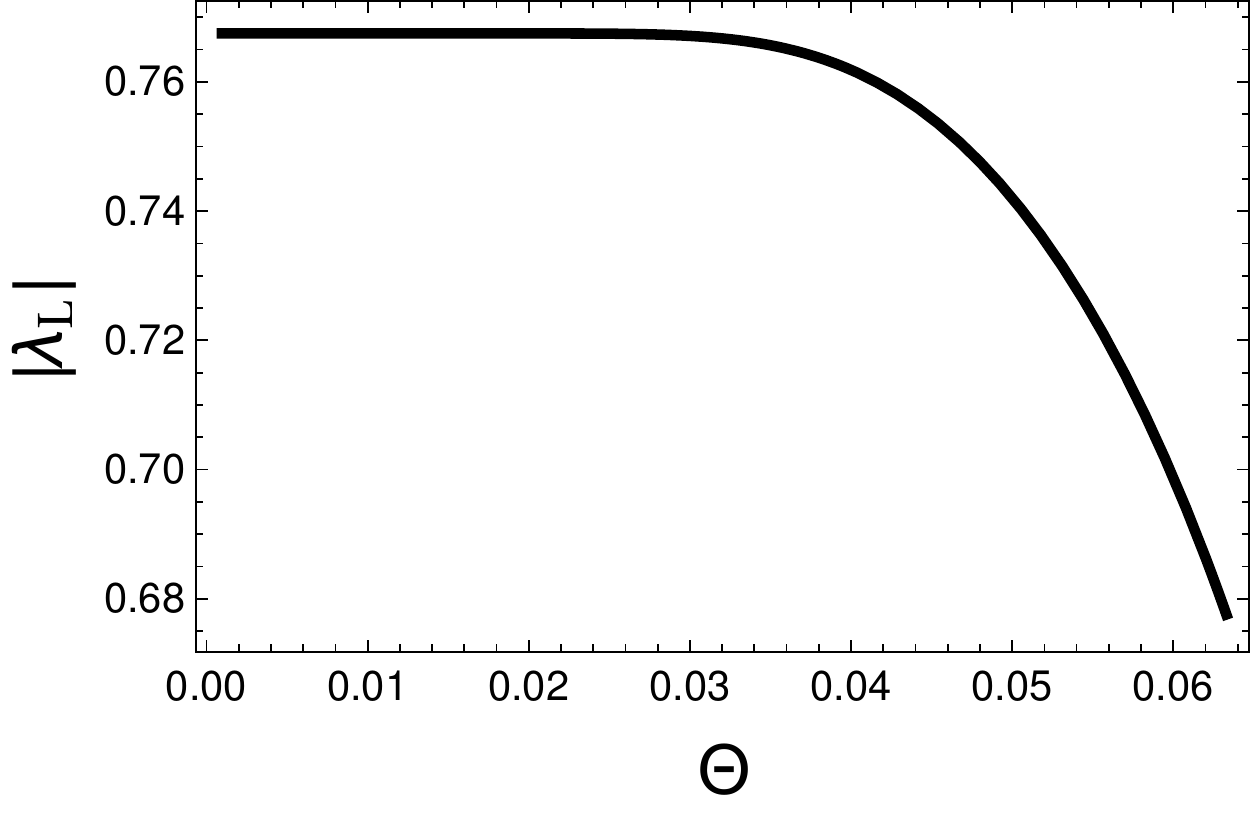}   \
\caption{
Critical frequency $\Omega_c$ (left panel) and Lyapunov exponent $\lambda_L$ (right panel) vs non-commutative parameter $\Theta$ for $M=1$. 
}
\label{fig:eikonal}
\end{figure*}

\section{QNMs of NC BHs in the WKB approximation}

\subsection{Numerical results}

Computing the QNMs of black holes analytically is possible only in some cases, see e.g. \cite{potential,ferrari,cardoso2,exact1,exact2,exact3,exact4,exact5,exact6,Ovgun:2018gwt,Rincon:2018ktz}. Semi-analytical methods based on the WKB approximation \cite{wkb1,wkb2,wkb3} are perhaps the most popular ones, and they have been applied extensively to several cases. For an incomplete list see e.g. \cite{paper1,paper2,paper3,paper4,paper5,paper6}, and for more recent works \cite{paper7,paper8,paper9,paper10,Rincon:2018sgd}, and references therein. 

The QN frequencies are given by
\begin{equation}
\omega^2 = V_0+(-2V_0'')^{1/2} \Lambda(n) - i \nu (-2V_0'')^{1/2} [1+\Omega(n)]
\end{equation}
where $n=0,1,2...$ is the overtone number, $\nu=n+1/2$, $V_0$ is the maximum of the effective potential, $V_0''$ is the second derivative of the effective potential evaluated at the maximum, while $\Lambda(n), \Omega(n)$ are complicated expressions of $\nu$ and higher derivatives of the potential evaluated at the maximum, and can be seen e.g. in \cite{paper2,paper7}. 

Here we have used the Wolfram Mathematica \cite{wolfram} code with WKB at any order from one to six (here we have worked in 6th order) presented in \cite{code} (see, however, \cite{Opala,Konoplya:2019hlu,RefExtra2} for higher order WKB corrections, and recipes for simple, quick, efficient and accurate computations). We have fixed the mass of the black hole to be $M=1$, the mass of the test scalar field is taken to be either $\mu=0$ or $\mu=0.15,0.3$, while for the noncommutative parameter $\Theta$ we have considered the range $0 < \Theta < 0.0634$. Since the WKB approximation works very well for $l > n$ \cite{roman}, here we shall consider the cases i) ${l=1,n=0}$, ii) ${l=2,n=0,n=1}$, iii) ${l=3,n=0,n=1,n=2}$
and iv) ${l=4,n=0,n=1,n=2,n=3}$. Finally, the eikonal regime $l \gg 1$ will be considered separately in the end before concluding our work.

Our numerical results for the QN modes of the NC regular black holes are summarized in the tables II,III and IV below separately for the 3 cases $\mu=0,0.15,0.3$. For comparison reasons the QNMs of the standard $(\Theta=0)$ 5D black hole are shown as well in Table V. For better visualization all modes corresponding to a given $\mu$ are shown in a single figure where the real part and the imaginary of the modes are put on the horizontal and vertical axes, respectively. All modes are complex numbers with a positive real part and a negative imaginary part. For a given mass $\mu$ and angular degree $l$, as the overtone number $n$ increases the real part of the modes decreases, while the absolute value of the imaginary part increases. Furthermore, for a given mass and overtone number, as the angular degree increases the real part increases as well while the absolute value of the imaginary part decreases. Finally, we see that the real part increases, while the absolute value of the imaginary part decreases with the mass of the scalar field.

\subsection{QNMs in the eikonal approximation}

Finally, in the eikonal approximation ($l \gg 1$) the WKB method becomes increasingly accurate, and it is possible to obtain analytical expressions for the QN frequencies. In the eikonal limit ($l \rightarrow \infty$) the angular momentum term is the dominant one in the effective potential
\begin{equation}
V(r) \approx \frac{f(r) l^2}{r^2} \equiv l^2 g(r)
\end{equation}
where we introduce a new function $g(r)=f(r)/r^2$, and it is easy to verify that the maximum of the potential is located at $r_1$ that is computed solving the following algebraic equation
\begin{equation}
2 f(r_1) - r_1 f'(r)|_{r_1} = 0
\end{equation}
Then, following the formalism developed in \cite{eikonal1}, the QN modes in the eikonal limit can be computed by the formula 
\begin{equation}
\omega_{l \gg 1} = \Omega_c l - i \left(n+\frac{1}{2}\right) |\lambda_L|
\end{equation}
where the Lyapunov exponent $\lambda_L$ is given by \cite{eikonal1}
\begin{equation}
\lambda_L = r_1^2 \sqrt{\frac{g''(r_1) g(r_1)}{2}}
\end{equation}
while the angular velocity $\Omega_c$ at the unstable null geodesic is given by \cite{eikonal1}
\begin{equation}
\Omega_c = \frac{\sqrt{f(r_1)}}{r_1}
\end{equation}
We see that the angular velocity determines the real part of the modes, where only the degree of angular momentum $l$ enters, while the Lyapunov exponent determines the imaginary part of the modes, where only the overtone number $n$ enters. In Fig. \eqref{fig:eikonal} we show the angular velocity (left panel) as well as the Lyapunov exponent (right panel) as a function of $\Theta$ for $M=1$.
The angular velocity increases monotonically with the noncommutative parameter, while the Lyapunov exponent decreases monotonically, similarly to the regular Bardeen black hole studied in \cite{paper7}, where it was found that $\lambda_L$ decreases monotonically and $\Omega_c$ increases monotonically with the charge $q$ of the black hole.

\begin{table*}
\centering
\caption{QN frequencies for $M=1, \Theta=0.04, \mu=0$.}
\begin{tabular}{ccccc}
\hline
$n$ & $l=1$ & $l=2$ & $l=3$ & $l=4$\\
0  & 0.539924 - 1.40113 i & 1.5548 - 0.424775 i  & 2.16755 - 0.3848 i  &  2.71583 - 0.381501 i \\
\hline
1  &                      & 1.0682 - 2.2103 i    & 1.93747 - 1.30872 i &  2.5945 - 1.18577 i   \\
\hline
2  &                      &                      & 1.54614 - 3.03226 i &  2.30746 - 2.24034 i  \\
\hline
3  &                      &                      &                     &  1.98674 - 3.94797 i  \\
\end{tabular}
\label{table:First_set}
\end{table*}

\begin{table*}
\centering
\caption{QN frequencies for $M=1, \Theta=0.04, \mu=0.15$.}
\begin{tabular}{ccccc}
\hline
$n$ & $l=1$               & $l=2$                 & $l=3$               & $l=4$           \\
0   & 0.534454 - 1.4136 i & 1.55935 - 0.422435 i  & 2.1702 - 0.384021 i &  2.71787 - 0.381066 i \\
\hline
1   &                     & 1.06649 - 2.20842 i   & 1.93937 - 1.30614 i &  2.59617 - 1.18448 i \\
\hline
2   &                     &                       & 1.54483 - 3.03194 i &  2.30792 - 2.23885 i \\
\hline
3   &                     &                       &                     &  1.98549 - 3.94925 i \\
\end{tabular}
\label{table:Second_set}
\end{table*}

\begin{table*}
\centering
\caption{QN frequencies for $M=1, \Theta=0.04, \mu=0.30$.}
\begin{tabular}{ccccc}
\hline
$n$ & $l=1$               & $l=2$                 & $l=3$               & $l=4$                 \\
0   & 0.519028 - 1.44467 i & 1.57336 - 0.41529 i  & 2.17818 - 0.381679 i &  2.72396 - 0.379759 i \\
\hline
1   &                      & 1.06181 - 2.20052 i   & 1.94521 - 1.29825 i &  2.6012 - 1.1806 i    \\
\hline
2   &                      &                       & 1.54097 - 3.03039 i &  2.30936 - 2.23431 i  \\
\hline 
3   &                      &                       &                     &  1.98177 - 3.95294 i  \\
\end{tabular}
\label{table:Third_set}
\end{table*}

\begin{table*}
\centering
\caption{QN frequencies for $M=1, \mu=0, \Theta=0$.}
\begin{tabular}{ccccc}
\hline
$n$ & $l=1$ & $l=2$ & $l=3$ & $l=4$\\
0  & 1.10108 -0.396434 i  & 1.6395 -0.388248 i  & 2.17936 -0.3862 i   &  2.72033 -0.385311 i \\
\hline
1  &                      & 1.51141 -1.19977 i  & 2.08063 -1.17802 i  &  2.64043 -1.16829 i  \\
\hline
2  &                      &                     & 1.89592 -2.02892 i &  2.48675 -1.98905 i  \\
\hline
3  &                      &                     &                     &  2.27294 -2.8726 i  \\
\end{tabular}
\label{table:Fourth_set}
\end{table*}

\section{Conclusions}\label{Conclusions}

In this article we have computed the quasinormal modes of five-dimensional black holes in the framework of noncommutative geometry. We have studied scalar perturbations using a Schr{\"o}dinger-like equation with the appropriate effective potential, and we have adopted the popular and extensively used WKB approximation of 6th order. All modes are found to be stable. Our numerical results are summarized in tables, and for better visualization we have shown graphically on the (real part-imaginary part) plane. For comparison reasons the QNMs of the standard 5D Schwarzschild black hole are shown as well.


\section*{Acknowlegements}

The author G.~P. thanks the Funda\c c\~ao para a Ci\^encia e Tecnologia (FCT), Portugal, for the financial support to the Center for Astrophysics and Gravitation-CENTRA,  Instituto Superior T\'ecnico,  Universidade de Lisboa, through the Grant No. UID/FIS/00099/2013.
The author \'A.~R. acknowledges DI-VRIEA for financial support through Proyecto Postdoctorado 2019 VRIEA-PUCV.



\begin{thebibliography}{99}
\bibitem{GR} A. Einstein, 
Annalen Phys. 49 (1916) 769–822.

\bibitem{Bardeen} J.~Bardeen, presented at GR5, Tiflis, U.S.S.R., and published 
in the conference proceedings in the U.S.S.R. (1968)

\bibitem{borde} A.~Borde,
  Phys.\ Rev.\ D {\bf 55} (1997) 7615
  [gr-qc/9612057].
  
\bibitem{beato1} E.~Ayon-Beato and A.~Garcia,
  Phys.\ Rev.\ Lett.\  {\bf 80} (1998) 5056
  [gr-qc/9911046].

\bibitem{beato2} E.~Ayon-Beato and A.~Garcia,
  Gen.\ Rel.\ Grav.\  {\bf 31} (1999) 629
  [gr-qc/9911084].

\bibitem{beato3} E.~Ayon-Beato and A.~Garcia,
  Phys.\ Lett.\ B {\bf 464} (1999) 25
  [hep-th/9911174].
  
\bibitem{bronnikov} K.~A.~Bronnikov,
  Phys.\ Rev.\ D {\bf 63} (2001) 044005
  [gr-qc/0006014].

\bibitem{dymnikova} I.~Dymnikova,
  Class.\ Quant.\ Grav.\  {\bf 21} (2004) 4417
  [gr-qc/0407072].

\bibitem{hayward} S.~A.~Hayward,
  Phys.\ Rev.\ Lett.\  {\bf 96} (2006) 031103
  [gr-qc/0506126].  

\bibitem{vagenas1} L.~Balart and E.~C.~Vagenas,
  Phys.\ Lett.\ B {\bf 730} (2014) 14
  [arXiv:1401.2136 [gr-qc]].

\bibitem{vagenas2} L.~Balart and E.~C.~Vagenas,
  Phys.\ Rev.\ D {\bf 90} (2014) no.12,  124045
  [arXiv:1408.0306 [gr-qc]].
  
\bibitem{RN} H.~Reissner, Annalen Phys. 355 (1916) 106-120.

\bibitem{ellis} S.~W.~Hawking and G.~F.~R.~Ellis, \textit{The Large Scale Structure of Space-Time}, Cambridge Monographs on Mathematical Physics (Cambridge University Press, Cambridge, England, 1973).

\bibitem{extra} J.~M.~M.~Senovilla,
  Gen.\ Rel.\ Grav.\  {\bf 30} (1998) 701
  [arXiv:1801.04912 [gr-qc]].
  
\bibitem{Connes1} A.~Connes, 
Publ. \ Math. \ IHES {\bf 62}, 44 (1983); \\
Noncommutative Geometry (Academic Press, New York 1994).
  
\bibitem{Connes2} A.~Connes and J.~Lott,
  Nucl.\ Phys.\ Proc.\ Suppl.\  {\bf 18B} (1991) 29.
  
\bibitem{Connes3} A.~Connes,
  J.\ Math.\ Phys.\  {\bf 41} (2000) 3832
  [hep-th/0003006]. 
  
\bibitem{Douglas} A.~Connes, M.~R.~Douglas and A.~S.~Schwarz,
  JHEP {\bf 9802} (1998) 003
  [hep-th/9711162].   

\bibitem{Seiberg} N.~Seiberg and E.~Witten,
  JHEP {\bf 9909} (1999) 032
  [hep-th/9908142].
  
\bibitem{ST1} M.~B.~Green, J.~H.~Schwarz and E.~Witten, \textit{Superstring Theory, Vol. 1 \& 2}, Cambridge Monographs on Mathematical Physics (Cambridge University Press, Cambridge, England, 2012).

\bibitem{ST2} J.~Polchinski, \textit{String Theory, Vol. 1 \& 2}, Cambridge Monographs on Mathematical Physics (Cambridge University Press, Cambridge, England, 2005).  
  
\bibitem{ADD} I.~Antoniadis, N.~Arkani-Hamed, S.~Dimopoulos and G.~R.~Dvali,
  Phys.\ Lett.\ B {\bf 436} (1998) 257
  [hep-ph/9804398].
  
\bibitem{RS1} L.~Randall and R.~Sundrum,
  Phys.\ Rev.\ Lett.\  {\bf 83} (1999) 3370
  [hep-ph/9905221].
  
\bibitem{RS2} L.~Randall and R.~Sundrum,
  Phys.\ Rev.\ Lett.\  {\bf 83} (1999) 4690
  [hep-th/9906064].

\bibitem{DGP} G.~R.~Dvali, G.~Gabadadze and M.~Porrati,
  Phys.\ Lett.\ B {\bf 485} (2000) 208
  [hep-th/0005016].

\bibitem{maldacena} J.~M.~Maldacena,
  Int.\ J.\ Theor.\ Phys.\  {\bf 38} (1999) 1113
   [Adv.\ Theor.\ Math.\ Phys.\  {\bf 2} (1998) 231].
   
\bibitem{klebanov} M.~K.~Benna and I.~R.~Klebanov,
  Les Houches {\bf 87} (2008) 611
  [arXiv:0803.1315 [hep-th]].
  
\bibitem{guide} M.~Natsuume,
  Lect.\ Notes Phys.\  {\bf 903} (2015) pp.1
  [arXiv:1409.3575 [hep-th]].
  
\bibitem{wheeler} T.~Regge and J.~A.~Wheeler,
  Phys.\ Rev.\  {\bf 108} (1957) 1063.

\bibitem{zerilli1} F.~J.~Zerilli,
  Phys.\ Rev.\ Lett.\  {\bf 24} (1970) 737.
  
\bibitem{zerilli2} F.~J.~Zerilli,
  Phys.\ Rev.\ D {\bf 2} (1970) 2141.
  
\bibitem{zerilli3} F.~J.~Zerilli,
  Phys.\ Rev.\ D {\bf 9} (1974) 860.

\bibitem{moncrief} V.~Moncrief,
  Phys.\ Rev.\ D {\bf 12} (1975) 1526.

\bibitem{teukolsky} S.~A.~Teukolsky,
  Phys.\ Rev.\ Lett.\  {\bf 29} (1972) 1114.

\bibitem{monograph} S.~Chandrasekhar, \textit{The mathematical theory of black holes}, OXFORD, UK: CLARENDON (1985) 646 P.
  
\bibitem{ligo1} B.~P.~Abbott {\it et al.} [LIGO Scientific and Virgo Collaborations],
  Phys.\ Rev.\ Lett.\  {\bf 116} (2016) no.6,  061102
[arXiv:1602.03837 [gr-qc]].

\bibitem{ligo2} B.~P.~Abbott {\it et al.} [LIGO Scientific and Virgo Collaborations],
  Phys.\ Rev.\ Lett.\  {\bf 116} (2016) no.24,  241103
[arXiv:1606.04855 [gr-qc]].

\bibitem{ligo3} B.~P.~Abbott {\it et al.} [LIGO Scientific and VIRGO Collaborations],
  Phys.\ Rev.\ Lett.\  {\bf 118} (2017) no.22,  221101
[arXiv:1706.01812 [gr-qc]].  

\bibitem{ligo4} B.~P.~Abbott {\it et al.} [LIGO Scientific and Virgo Collaborations],
  Phys.\ Rev.\ Lett.\  {\bf 119} (2017) no.14,  141101
  [arXiv:1709.09660 [gr-qc]].

\bibitem{ligo5} B.~P.~Abbott {\it et al.} [LIGO Scientific and Virgo Collaborations],
  Astrophys.\ J.\  {\bf 851} (2017) no.2,  L35
  [arXiv:1711.05578 [astro-ph.HE]].

\bibitem{pani} V.~Cardoso, S.~Hopper, C.~F.~B.~Macedo, C.~Palenzuela and P.~Pani,
  Phys.\ Rev.\ D {\bf 94} (2016) no.8,  084031
  [arXiv:1608.08637 [gr-qc]].

\bibitem{review1} K.~D.~Kokkotas and B.~G.~Schmidt,
  Living Rev.\ Rel.\  {\bf 2} (1999) 2
[gr-qc/9909058].

\bibitem{review2} E.~Berti, V.~Cardoso and A.~O.~Starinets,
  Class.\ Quant.\ Grav.\  {\bf 26} (2009) 163001
  [arXiv:0905.2975 [gr-qc]].
  
\bibitem{review3} R.~A.~Konoplya and A.~Zhidenko,
  Rev.\ Mod.\ Phys.\  {\bf 83} (2011) 793
  [arXiv:1102.4014 [gr-qc]].
  
\bibitem{extra1} P.~Bizon, T.~Chmaj and B.~G.~Schmidt,
  Phys.\ Rev.\ Lett.\  {\bf 95} (2005) 071102
  [gr-qc/0506074].

\bibitem{extra2} P.~Bizon, T.~Chmaj, A.~Rostworowski, B.~G.~Schmidt and Z.~Tabor,
  Phys.\ Rev.\ D {\bf 72} (2005) 121502
  [gr-qc/0511064].

\bibitem{kanti} P.~Kanti,
  Int.\ J.\ Mod.\ Phys.\ A {\bf 19} (2004) 4899
  [hep-ph/0402168].

\bibitem{extra3} S.~Hod,
  Phys.\ Rev.\ Lett.\  {\bf 81} (1998) 4293
  [gr-qc/9812002].

\bibitem{extra4} G.~Kunstatter,
  Phys.\ Rev.\ Lett.\  {\bf 90} (2003) 161301
  [gr-qc/0212014].
  
\bibitem{Lemos_New} A.~Flachi and J.~P.~S.~Lemos,
  Phys.\ Rev.\ D {\bf 87} (2013) no.2,  024034
  [arXiv:1211.6212 [gr-qc]].
  
\bibitem{GPAR} G.~Panotopoulos and \'A.~Rinc{\'o}n,
  Eur.\ Phys.\ J.\ Plus {\bf 134} (2019) no.6,  300
  [arXiv:1904.10847 [gr-qc]].
  
\bibitem{churilova} M.~S.~Churilova and Z.~Stuchlik,
  arXiv:1910.12660 [gr-qc].  
  
\bibitem{Kobakhidze:2016cqh} A.~Kobakhidze, C.~Lagger and A.~Manning,
  Phys.\ Rev.\ D {\bf 94}, no. 6, 064033 (2016)
  [arXiv:1607.03776 [gr-qc]].  
    
\bibitem{work1} K.~S.~Gupta, E.~Harikumar, T.~Jurić, S.~Meljanac and A.~Samsarov,
  JHEP {\bf 1509} (2015) 025
  [arXiv:1505.04068 [hep-th]].
  
\bibitem{work2} K.~S.~Gupta, T.~Jurić and A.~Samsarov,
  JHEP {\bf 1706} (2017) 107
  [arXiv:1703.00514 [hep-th]].
  
\bibitem{work3} J.~Liang,
  Chin.\ Phys.\ Lett.\  {\bf 35} (2018) no.1,  010401.

\bibitem{work4} J.~Liang,
  Chin.\ Phys.\ Lett.\  {\bf 35} (2018) no.5,  050401.

\bibitem{work5} M.~D.~{\'C}iri{\'c}, N.~Konjik and A.~Samsarov,
  Class.\ Quant.\ Grav.\  {\bf 35} (2018) no.17,  175005
  [arXiv:1708.04066 [hep-th]].
  
\bibitem{RefExtra1} K.~Das, S.~Pramanik and S.~Ghosh,
  Phys.\ Rev.\ D {\bf 99} (2019) no.2,  024039
  [arXiv:1807.08517 [hep-th]].
  
\bibitem{5D1} V.~Cardoso, O.~J.~C.~Dias and J.~P.~S.~Lemos,
  Phys.\ Rev.\ D {\bf 67} (2003) 064026
  [hep-th/0212168].

\bibitem{5D2} R.~A.~Konoplya,
  Phys.\ Rev.\ D {\bf 68} (2003) 024018
  [gr-qc/0303052].
    
\bibitem{molina} C.~Molina,
  Phys.\ Rev.\ D {\bf 68} (2003) 064007
  [gr-qc/0304053].
  
\bibitem{master} H.~Kodama and A.~Ishibashi,
  Prog.\ Theor.\ Phys.\  {\bf 110} (2003) 701
  [hep-th/0305147].
  
\bibitem{5D3} R.~A.~Konoplya,
  Phys.\ Rev.\ D {\bf 68} (2003) 124017
  [hep-th/0309030].
  
\bibitem{5D4} V.~Cardoso, J.~P.~S.~Lemos and S.~Yoshida,
  Phys.\ Rev.\ D {\bf 69} (2004) 044004
  [gr-qc/0309112].

\bibitem{5D5} E.~Berti, M.~Cavaglia and L.~Gualtieri,
  Phys.\ Rev.\ D {\bf 69} (2004) 124011
  [hep-th/0309203].
 
\bibitem{ortega1} A.~Lopez-Ortega,
  Gen.\ Rel.\ Grav.\  {\bf 38} (2006) 1565
  [gr-qc/0605027].
  
\bibitem{ortega2} A.~Lopez-Ortega,
  Gen.\ Rel.\ Grav.\  {\bf 38} (2006) 1747
  [gr-qc/0605034].
  
\bibitem{ortega3} A.~Lopez-Ortega,
  Gen.\ Rel.\ Grav.\  {\bf 39} (2007) 1011
  [arXiv:0704.2468 [gr-qc]].
  
\bibitem{ortega4} A.~Lopez-Ortega,
  Gen.\ Rel.\ Grav.\  {\bf 40} (2008) 1379
  [arXiv:0706.2933 [gr-qc]].
  
\bibitem{panotopoulos} G.~Panotopoulos,
  Mod.\ Phys.\ Lett.\ A {\bf 33} (2018) no.23,  1850130
  [arXiv:1807.03278 [gr-qc]].
  
\bibitem{Smailagic:2003yb} 
  A.~Smailagic and E.~Spallucci,
  J.\ Phys.\ A {\bf 36}, L467 (2003)
  [hep-th/0307217].

\bibitem{NCBH1} P.~Nicolini, A.~Smailagic and E.~Spallucci,
  Phys.\ Lett.\ B {\bf 632} (2006) 547
  [gr-qc/0510112].

\bibitem{NCBH2} Y.~M.~Wu and Y.~G.~Miao,
  arXiv:1810.08984 [hep-th].
  
\bibitem{Ivan} I.~Arraut, D.~Batic and M.~Nowakowski,
  J.\ Math.\ Phys.\  {\bf 51} (2010) 022503
  [arXiv:1001.2226 [gr-qc]].
  
\bibitem{EGB1} K.~Zhou, Z.~Y.~Yang, D.~C.~Zou and R.~H.~Yue,
  Chin.\ Phys.\ B {\bf 21} (2012) 020401
  [arXiv:1107.2732 [gr-qc]].

\bibitem{EGB2} S.~Hansraj, B.~Chilambwe and S.~D.~Maharaj,
  Eur.\ Phys.\ J.\ C {\bf 75} (2015) no.6,  277
  [arXiv:1502.02219 [gr-qc]].
  
\bibitem{Tangherlini} F.~R.~Tangherlini,
  Nuovo Cim.\  {\bf 27} (1963) 636.
  
\bibitem{Moyal} J.~E.~Moyal,
  Proc.\ Cambridge Phil.\ Soc.\  {\bf 45} (1949) 99.

\bibitem{lectures} F.~A.~Schaposnik,
  hep-th/0408132.  

\bibitem{book} C.~Muller, in \textit{Lecture Notes in Mathematics: Spherical Harmonics} (Springer-Verlag, Berlin-Heidelberg, 1966).
  
\bibitem{valeria} V.~Ferrari and L.~Gualtieri,
  Gen.\ Rel.\ Grav.\  {\bf 40} (2008) 945
  [arXiv:0709.0657 [gr-qc]].  

\bibitem{potential} G.~Poschl and E.~Teller,
  Z.\ Phys.\  {\bf 83} (1933) 143.
  
\bibitem{ferrari} V.~Ferrari and B.~Mashhoon,
  Phys.\ Rev.\ D {\bf 30} (1984) 295.  

\bibitem{cardoso2} V.~Cardoso and J.~P.~S.~Lemos,
  Phys.\ Rev.\ D {\bf 63} (2001) 124015
[gr-qc/0101052].

\bibitem{exact1} D.~Birmingham,
  Phys.\ Rev.\ D {\bf 64} (2001) 064024
[hep-th/0101194].

\bibitem{exact2} V.~Cardoso and J.~P.~S.~Lemos,
  Phys.\ Rev.\ D {\bf 67} (2003) 084020
  doi:10.1103/PhysRevD.67.084020
  [gr-qc/0301078].  
  
\bibitem{exact3} C.~Molina,
  Phys.\ Rev.\ D {\bf 68} (2003) 064007
  [gr-qc/0304053].

\bibitem{exact4} S.~Fernando,
  Gen.\ Rel.\ Grav.\  {\bf 36} (2004) 71
  [hep-th/0306214].

\bibitem{exact5} S.~Fernando,
  Phys.\ Rev.\ D {\bf 77} (2008) 124005
  [arXiv:0802.3321 [hep-th]].

\bibitem{exact6} K.~Destounis, G.~Panotopoulos and \'A.~Rinc\'on,
  Eur.\ Phys.\ J.\ C {\bf 78} (2018) no.2,  139
  [arXiv:1801.08955 [gr-qc]].
  
\bibitem{Ovgun:2018gwt} A.~Ovg\"un and K.~Jusufi,
  Annals Phys.\  {\bf 395}, 138 (2018)
  [arXiv:1801.02555 [gr-qc]].  
  
\bibitem{Rincon:2018ktz} \'A.~Rinc\'on and G.~Panotopoulos,
  Eur.\ Phys.\ J.\ C {\bf 78}, no. 10, 858 (2018)
  [arXiv:1810.08822 [gr-qc]].
  
\bibitem{wkb1} B.~F.~Schutz and C.~M.~Will,
  Astrophys.\ J.\  {\bf 291} (1985) L33.

\bibitem{wkb2} S.~Iyer and C.~M.~Will,
  Phys.\ Rev.\ D {\bf 35} (1987) 3621.

\bibitem{wkb3} R.~A.~Konoplya,
  Phys.\ Rev.\ D {\bf 68} (2003) 024018
[gr-qc/0303052].

\bibitem{paper1} S.~Iyer,
  Phys.\ Rev.\ D {\bf 35} (1987) 3632.

\bibitem{paper2} K.~D.~Kokkotas and B.~F.~Schutz,
  Phys.\ Rev.\ D {\bf 37} (1988) 3378.

\bibitem{paper3} E.~Seidel and S.~Iyer,
  Phys.\ Rev.\ D {\bf 41} (1990) 374.
  
\bibitem{paper4} R.~Konoplya,
  Phys.\ Rev.\ D {\bf 71} (2005) 024038
[hep-th/0410057].

\bibitem{paper5} S.~Fernando and C.~Holbrook,
  Int.\ J.\ Theor.\ Phys.\  {\bf 45} (2006) 1630
  [hep-th/0501138].
  
\bibitem{paper6} S.~K.~Chakrabarti,
  Gen.\ Rel.\ Grav.\  {\bf 39} (2007) 567
[hep-th/0603123].  

\bibitem{paper7} S.~Fernando and J.~Correa,
  Phys.\ Rev.\ D {\bf 86} (2012) 064039
  [arXiv:1208.5442 [gr-qc]].

\bibitem{paper8} V.~Santos, R.~V.~Maluf and C.~A.~S.~Almeida,
  Phys.\ Rev.\ D {\bf 93} (2016) no.8,  084047
[arXiv:1509.04306 [gr-qc]].

\bibitem{paper9} J.~L.~Bl{\'a}zquez-Salcedo, F.~S.~Khoo and J.~Kunz,
  arXiv:1706.03262 [gr-qc].
  
\bibitem{paper10} G.~Panotopoulos and \'A.~Rinc{\'o}n,
  Int.\ J.\ Mod.\ Phys.\ D {\bf 27} (2017) no.03,  1850034
  [arXiv:1711.04146 [hep-th]].  

\bibitem{Rincon:2018sgd} \'A.~Rinc\'on and G.~Panotopoulos,
  Phys.\ Rev.\ D {\bf 97}, no. 2, 024027 (2018)
  [arXiv:1801.03248 [hep-th]].
  
\bibitem{wolfram} 
\url{http://www.wolfram.com}

\bibitem{code} R.~A.~Konoplya and A.~Zhidenko,
  Phys.\ Rev.\ D {\bf 81} (2010) 124036
  [arXiv:1004.1284 [hep-th]].
  
\bibitem{Opala} J.~Matyjasek and M.~Opala,
  Phys.\ Rev.\ D {\bf 96} (2017) no.2,  024011
  [arXiv:1704.00361 [gr-qc]].

\bibitem{Konoplya:2019hlu} R.~A.~Konoplya, A.~Zhidenko and A.~F.~Zinhailo,
  Class.\ Quant.\ Grav.\  {\bf 36} (2019) 155002
  [arXiv:1904.10333 [gr-qc]].  
  
\bibitem{RefExtra2} Y.~Hatsuda,
  arXiv:1906.07232 [gr-qc].
  
\bibitem{roman} R.~Konoplya,
  Phys.\ Rev.\ D {\bf 71} (2005) 024038
  [hep-th/0410057].  

\bibitem{eikonal1} V.~Cardoso, A.~S.~Miranda, E.~Berti, H.~Witek and V.~T.~Zanchin,
  Phys.\ Rev.\ D {\bf 79} (2009) 064016
  [arXiv:0812.1806 [hep-th]].
\end{thebibliography}
\end{document}